%% file: conference_101719.tex
\documentclass[conference]{IEEEtran}
\IEEEoverridecommandlockouts
\usepackage{amsmath,amssymb,amsfonts}
\usepackage{algorithmic}
\usepackage{graphicx}
\usepackage{textcomp}
\usepackage{xcolor}
\usepackage{subfig}
\usepackage{hyperref}
\usepackage{xspace}

\newcommand{\supremica}{\textsc{Supremica}\xspace}

\begin{document}

\title{Identification of Minimally Restrictive Assembly Sequences using Supervisory Control Theory
\thanks{This work was supported by the EUREKA ITEA4 ArtWork project (2023-00970), and the Wallenberg AI, Autonomous Systems and Software Program (WASP) funded by the Knut and Alice Wallenberg Foundation.}
}

\author{\IEEEauthorblockN{Martina Vinetti}
\IEEEauthorblockA{\textit{Department of Electrical Engineering} \\
\textit{Chalmers University of Technology}\\
G{\"o}teborg, Sweden \\
vinetti@chalmers.se}
\and
\IEEEauthorblockN{Martin Fabian}
\IEEEauthorblockA{\textit{Department of Electrical Engineering} \\
\textit{Chalmers University of Technology}\\
G{\"o}teborg, Sweden \\
fabian@chalmers.se}
}

\maketitle

\begin{abstract}
Modern assembly processes require flexibility and adaptability to handle increasing product variety and customization. Traditional assembly planning methods often prioritize finding an optimal assembly sequence, overlooking the requirements of contemporary manufacturing. This work uses \textit{Supervisory Control Theory} to systematically generate \emph{all} feasible assembly sequences while ensuring compliance with precedence and process constraints. By synthesizing a controllable, non-blocking, and minimally restrictive supervisor,
%%MF in \textsc{Supremica}, 
the proposed method guarantees that only valid sequences are allowed, balancing flexibility and constraint enforcement. The obtained sequences can serve as a basis for further optimization or exception management, improving responsiveness to disruptions.
\end{abstract}

%\begin{IEEEkeywords}
%component, formatting, style, styling, insert
%\end{IEEEkeywords}

\section{Introduction}
\input{Introduction}

\section{Supervisory Control Theory}
\input{SCT}

\section{Method}
\input{Method}

\section{Case Study}
\input{CaseStudy}

\section{Conclusions}
\input{Conclusions}

\bibliographystyle{IEEEtran}
\bibliography{References}

\end{document}

%% file: Introduction.tex
In recent years, assembly processes have encountered increasing challenges due to evolving market demands and technological advancements~\cite{assemblychallenges}. The explosion in product variants~\cite{productvariant} and the growing emphasis on customization~\cite{masscustomiz} have added complexity to assembly lines, requiring greater flexibility and adaptability while maintaining efficiency and quality.

%Thus, flexibility and adaptive assembly sequence planning have become critical needs in modern manufacturing systems. In industries such as automotive~\cite{flexibleassembly}, aerospace, and consumer electronics, where product diversity is most pronounced, it is vital that assembly processes quickly reconfigure for new designs or variations.

These challenges are especially critical in manual assembly, where workers must manage increasing product variations and adapt task sequences to specifications, disruptions, or resource availability. This underscores the need for flexible and automated support in assembly sequence planning to ensure efficiency while reducing cognitive overload.

Traditional assembly planning methods often emphasize the search for a single optimal sequence, a challenge known as Assembly Sequence Planning (ASP)~\cite{asp}, but may struggle to meet the demands of modern manufacturing. These approaches prioritize optimization metrics, such as cost, time, and efficiency, over the flexibility and adaptability needed to handle dynamic and uncertain production environments. For example,~\cite{aspref1} introduces a hybrid Symbiotic Organisms search and Ant Colony Optimization algorithm (SOS-ACO), optimizing costs and efficiency while adhering to assembly constraints. Similarly,~\cite{aspref2} proposes a Discrete Particle Swarm Optimization (DPSO) algorithm to minimize assembly costs and time.

This work addresses the challenge of generating \emph{all} feasible assembly sequences that meet specified constraints. 
%%MF by utilizing Supervisory Control Theory (SCT). 
%%MF Unlike conventional methods, which often rely on specific representations like interference matrices and precedence graphs tailored to particular constraints, SCT provides a unified and flexible modeling framework. 
This is achieved by using the Supervisory Control Theory (SCT) which, in contrast to conventional methods that often rely on specific representations like interference matrices~\cite{aspref1} and precedence graphs tailored to particular constraints~\cite{asp}, provides a unified and flexible modeling framework. 
Assembly tasks and the constraints between them are represented as finite automata (FAs), enabling the seamless incorporation of both static and dynamic constraints while maintaining a comprehensive perspective of the assembly process. Additionally, this method avoids blocking configurations, from where no feasible sequence of assembly tasks can lead to completion without violating some constraint. 
%%MF If a blocking configuration is detected, the events that led to it will prevented. 

The minimally restrictive nature of the synthesized supervisor, an inherent property of SCT, ensures that assembly tasks are restricted only when necessary, balancing feasibility and flexibility. This guarantees that any sequence obtained allows the completion of the entire assembly process while satisfying all constraints, addressing the need for adaptive and robust assembly planning in complex and dynamic manufacturing environments.

The generation of all feasible assembly sequences offers significant advantages. It confines the search space for further optimization, making it easier to identify the most suitable sequence according to specific performance criteria. Also, it provides a foundation for exception management systems, enabling the identification of alternative sequences when disruptions occur, such as missing components or unavailable tools. Moreover, in manual assembly,  by giving operators various feasible sequences, adaptability on the shop floor is improved, enabling workers to select the sequence that best suits their current context.  This aligns with the principles of Industry 5.0~\cite{industry5.0}, fostering a human-centric manufacturing environment where collaboration between advanced technologies and human operators enhances flexibility, personalization, and overall efficiency.

%This work aims to contribute to the development of more flexible and robust assembly planning frameworks. By leveraging SCT to model a wide range of constraints and task interactions, the proposed method advances the state of the art in assembly sequence generation and provides a foundation for future optimization and exception management strategies.

%% file: SCT.tex
A discrete event system can be modeled using finite automata (FAs)~\cite{des}. An FA is defined as a 5-tuple 
\(\langle Q, \Sigma, \delta, q_i, Q_m \rangle\), where \(Q\) represents the finite set of states, \(\Sigma\) is the finite 
set of events, \(\delta: Q \times \Sigma \to Q\) is the partial transition function, \(q_i \in Q\) is 
the initial state, and \(Q_m\) is the set of marked states.

In Supervisory Control Theory (SCT)~\cite{sct}, a system's behavior is modeled by an automaton known as the plant \(P\), which represents all possible behaviors of the system. To ensure that the system meets specific desired outcomes, the plant is restricted by the specification \(K\), another automaton that defines 
%%MF acceptable and prohibited states, events, and transitions.
the desired behavior.

When multiple FAs interact, their combined behavior can be represented by the \textit{synchronous composition} operator. This operator models the concurrent execution of two FAs by ensuring that shared events occur simultaneously in both systems while independent events evolve freely in their respective FA. Formally, given two FAs $A = (Q_A, \Sigma_A, \delta_A, q_i^A, Q_m^A)$ and $B = (Q_B, \Sigma_B, \delta_B, q_i^B, Q_m^B)$, their synchronous composition is defined as the FA:
\[
A \parallel B =
\langle Q_A \times Q_B, \Sigma_A \cup \Sigma_B, \delta, (q_i^A, q_i^B), Q_m^A \times Q_m^B \rangle
\]
where 
%%MF $Q \subseteq Q_A \times Q_B$ and $Q_m \subseteq Q_m^A \times Q_m^B$ and 
the transition function $\delta$ is given by:
\[
\delta((q^A, q^B), \sigma) =
\begin{cases} 
(\delta_A(q^A, \sigma), \delta_B(q^B, \sigma)) & \text{if } \sigma \in \Sigma_A \cap \Sigma_B \\
(\delta_A(q^A, \sigma), q^B) & \text{if } \sigma \in \Sigma_A \setminus \Sigma_B \\
(q^A, \delta_B(q^B, \sigma)) & \text{if } \sigma \in \Sigma_B \setminus \Sigma_A
\end{cases}
\]

%%MF By composing \(P\) and \(K\), a supervisor~\cite{sct} can be synthesized to control $P$’s behavior. The supervisor determines which events are permissible and ensures that marked states—typically representing task completion—remain reachable. Since some events are uncontrollable, the supervisor must enforce constraints while guaranteeing that a marked state is always accessible from any reachable state, ensuring non-blocking behavior. Furthermore, the supervisor should be minimally restrictive, imposing only the necessary constraints to satisfy \(K\).

From the composition $P||K$, a supervisor~\cite{sct} $S$ can be synthesized such that the controlled system $P||S$ fulfills $P||K$. Some events are \emph{uncontrollable} and thus not subject to influence by $S$, which must be taken into account by the synthesis. Also, $S$ must enforce the specification while guaranteeing that a marked state is always accessible from any reachable state, thus ensuring \emph{non-blocking} behavior. Lastly, $S$ should be \emph{minimally restrictive}, meaning that it does not impose more control than absolutely necessary to fulfill $K$.

The software tool \supremica~\cite{supremica} facilitates the synthesis of supervisors and provides formal verification to ensure that these supervisors satisfy the requirements of controllability, non-blocking, and minimal restrictiveness. \textsc{Supremica} has been applied in various domains, including industrial research projects related to robotics and manufacturing systems~\cite{exsupr}.  

%% file: Method.tex
Identifying feasible assembly sequences requires a precise understanding of the tasks involved in the assembly process and the constraints governing their execution. 
%This study focuses on a single-worker assembly process. However, the approach could be extended to a multi-worker scenario if additional considerations are taken to account for the parallel execution of tasks and the dependencies between tasks assigned to different workers.In the single-worker case analyzed here, the set of tasks assigned to the worker and the constraints between them are assumed to be known. 
This study considers a scenario where a set of tasks and their constraints are given, without addressing coordination between multiple agents (robots or workers). However, the approach can be easily extended to multi-agent settings by incorporating additional constraints on inter-agent task dependencies.
In the considered case, constraints are categorized as either \emph{static} or \emph{dynamic}, both imposing limitations on the possible assembly sequences.

Static constraints, such as precedence constraints, impose fixed limitations on the order of execution. These arise primarily from product design requirements, ensuring that parts are assembled in a way that satisfies technical and functional specifications. Additionally, some precedence constraints are dictated by process dependencies, which are not derivable solely from the product’s geometry but stem from operational requirements or interconnections with other components.

Beyond static constraints, real-time conditions may impose dynamic constraints.
Examples include tool/resource dependencies, where one task provides essential tools or resources for another; collision-free path dependencies, which ensure that assembly movements do not interfere with each other; and stability dependencies, where the structural stability of a subassembly depends on another ongoing operation.

To systematically represent both tasks and constraints, a formal modeling approach based on FAs in \supremica is adopted. The following section details how these elements are encoded within this framework.

\subsection{Modeling}
Each task in the assembly process is modeled as an individual finite automaton. Since, in a real system, only the initiation of a task can be controlled while its completion depends on uncontrollable factors, each task automaton has a controllable event representing the task's start and an uncontrollable event indicating its completion. As a result, tasks can be initiated or completed while others are executing, enabling the representation of process dependencies such as tool availability, collision-free path generation, or subassembly stability.

%This modeling approach does not restrict task execution sequences, allowing multiple tasks to be in progress simultaneously.

\begin{figure}[h]
    \centering
    \subfloat[Single-execution task modeling]{\includegraphics[width=0.15\textwidth]{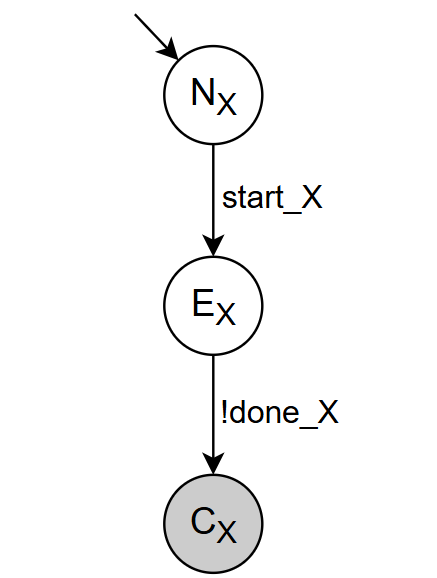}}
    \label{fig:a}
    \hspace{0.5cm}
    \subfloat[Repetitive task modeling]{\includegraphics[width=0.2\textwidth]{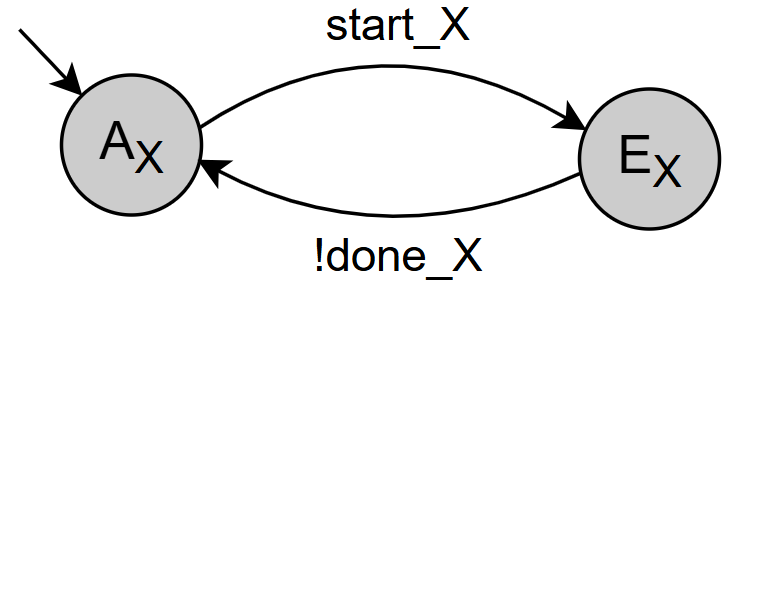}}
    \label{fig:b}
    \caption{FAs of a generic task X: (a) single-execution task with states \textit{Non-completed}, \textit{Executing}, and \textit{Completed}; (b) repetitive task with states \textit{Available} and \textit{Executing}.}
    \label{fig:tasksmodeling}
\end{figure}

Two types of tasks are considered, see Figure~\ref{fig:tasksmodeling}. Tasks that 
are executed only once during the assembly process are modeled as three-state automata, where the execution follows a finite sequence of transitions, ensuring a strict progression from initiation to completion. The final state is marked, as it represents the desired completion of the task. Conversely, tasks that require multiple executions, such as screwing operations or visual inspections, are represented using loop automata, with both states marked, allowing their execution to be repeated as needed within the process.
\vspace{-1em}

Both tasks and constraints can be modeled in multiple ways, based on states and events.

\subsection{Supervisor Synthesis}
Once the plant (the tasks) and the specifications (the constraints) have been modeled and synchronized, a supervisor~\cite{sct}
is synthesized in \supremica.
%%MF , using the monolithic approach. In this process of supervisor synthesis, an iterative procedure
Supervisor synthesis is an iterative procedure that 
eliminates undesirable states, specifically those from which no marked state is reachable or where uncontrollable events could lead to such states.
%%MF As has been demonstrated~\cite{sct}, this refinement process 
As is known~\cite{sct}, this procedure
converges to a fixpoint, yielding the least restrictive supervisor that ensures both controllability and nonblocking behavior. The supervisor guarantees that all allowed task sequences comply with the given constraints while maximizing system flexibility and ensuring that some marked state is always reachable.

%% file: CaseStudy.tex
\begin{figure}[h]
    \centering
    \includegraphics[width=0.15\textwidth]{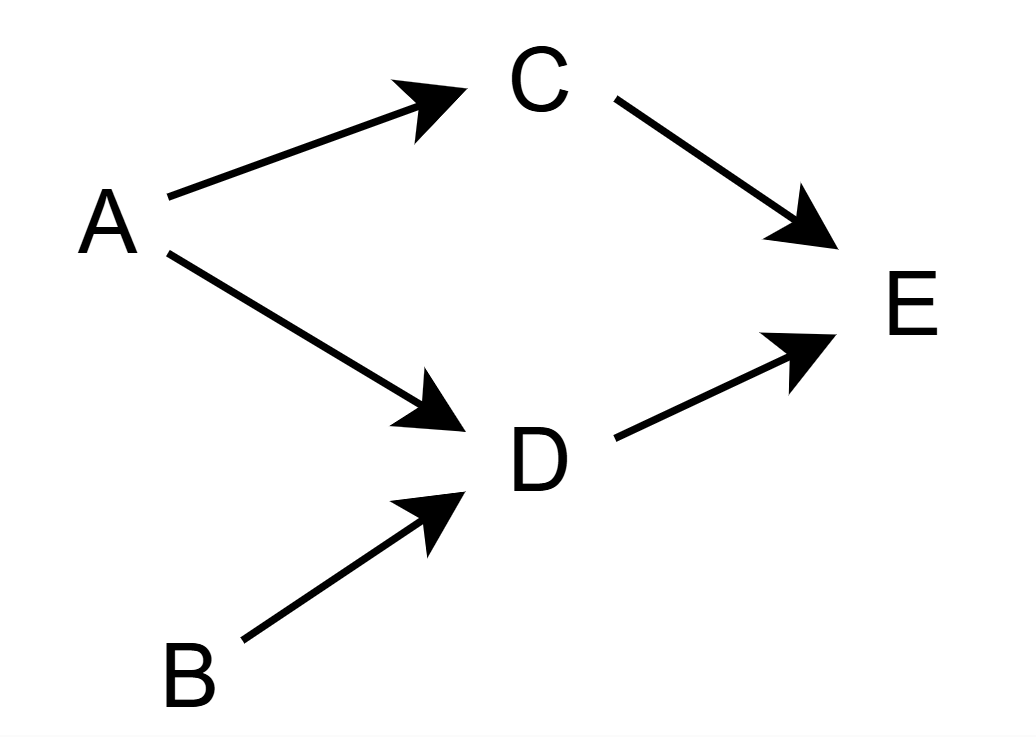}
    \caption{Digraph of precedence constraints among tasks.}
    \label{fig:precedence}
\end{figure}

A manageable case study has been selected, which allows for the manual verification of the expected behavior. It regards an assembly process consisting of five main tasks: \textit{A}, \textit{B}, \textit{C}, \textit{D}, and \textit{E}. Each task must be executed exactly once for the assembly process to be considered complete. The precedence constraints among these tasks are represented by the directed graph (\textit{digraph}) in Figure~\ref{fig:precedence}, where nodes correspond to tasks and directed edges indicate that a task must be done before the next can start.\vspace{-1.4em}

In \supremica, these tasks are represented as illustrated in Figure~\ref{fig:tasksmodeling}\,(a). In addition to static precedence constraints, dynamic constraints are also introduced.

\begin{figure}[h]
    \centering
    \includegraphics[width=0.25\textwidth]{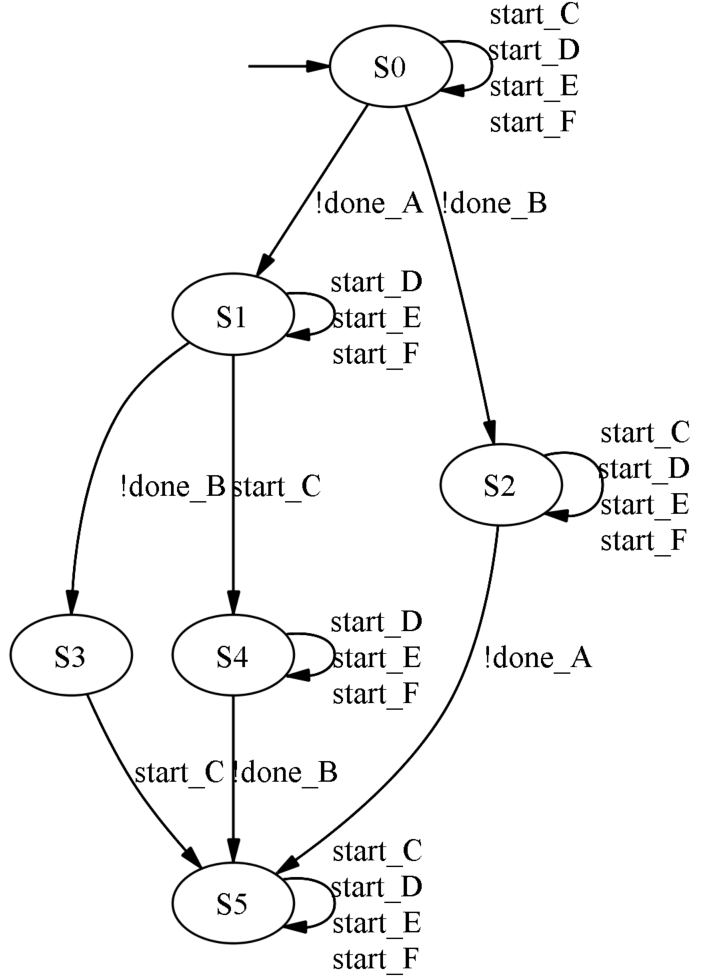}
    \caption{FA representation of the second dynamic constraint.}
    \label{fig:dyncon2}
\end{figure}

Two dynamic constraints are considered:
\begin{enumerate}
    \item Task \textit{C} cannot start if task \textit{D} is done. In Figure~\ref{fig:dyncon1}, a possible FA-based model is shown for this constraint.\vspace{-1.4em}

    \begin{figure}[h]
        \centering
        \includegraphics[width=0.15\textwidth]{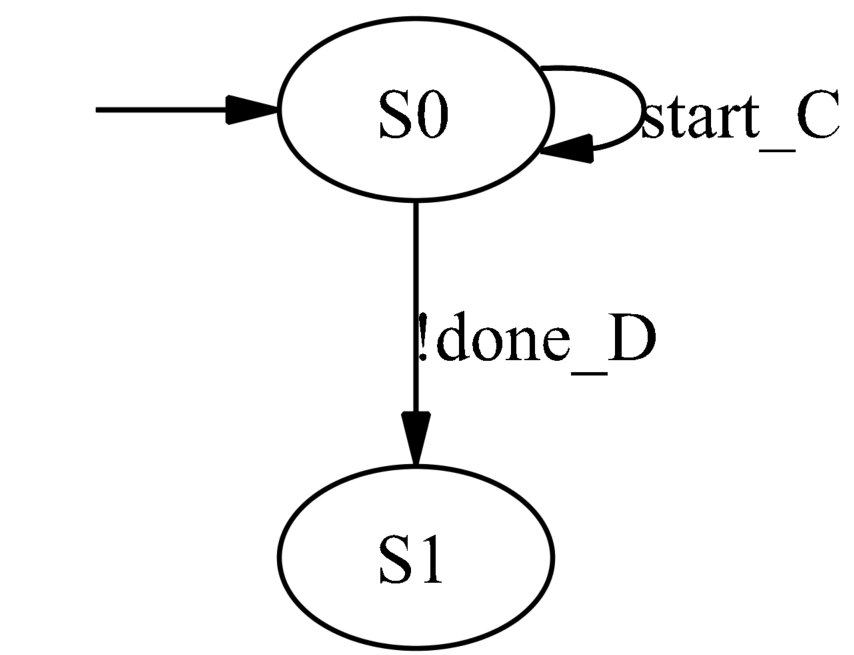}
        \caption{FA representation of the first dynamic constraint.}
        \label{fig:dyncon1}
    \end{figure}
    
    \iffalse
    \begin{equation}
        d_D \rightarrow \neg s_C
    \end{equation}
    \fi
    
    \item 
    If \textit{A} is done before \textit{B} and \textit{C} has not yet started, then \textit{C} should start immediately after \textit{B} is done (Figure~\ref{fig:dyncon2}).
    %task \textit{C} must be the next task to start (Figure~\ref{fig:dyncon2}).

    %If \textit{A} is completed before \textit{B}, then immediately after completing \textit{B}, task \textit{C} must start. 
    %Defining a Boolean variable $p_\mathit{AB}$, which is set to 1 if \textit{A} is completed before \textit{B}, this constraint is expressed as:
    \iffalse
    \begin{equation}
        p_\mathit{AB} \rightarrow (d_B \rightarrow s_C)
    \end{equation}
    \fi
\end{enumerate}

In this context, the term \textit{immediately} denotes a strict precedence relation, ensuring no other tasks start or are done between the specified ones.

In addition to these five assembly tasks, the process includes a repeatable task, \textit{F}, representing a \textit{screwing operation}. This task is modeled as a loop automaton, as shown in Figure~\ref{fig:tasksmodeling}\,(b), and its repeatability is governed by a dynamic constraint consisting of two parts in sequence. The first part is an \textit{if-clause}: if task $A$ is done immediately after $B$ is done, then $F$ must start immediately and be done before $C$ and $D$ start. The second part applies unconditionally: $F$ must start immediately after $C$ and $D$ are done and must be done before $E$ starts.  Due to space limitations, the FA model of this constraint is not shown.

\begin{figure}[h]
    \centering
    \includegraphics[width=0.25\textwidth]{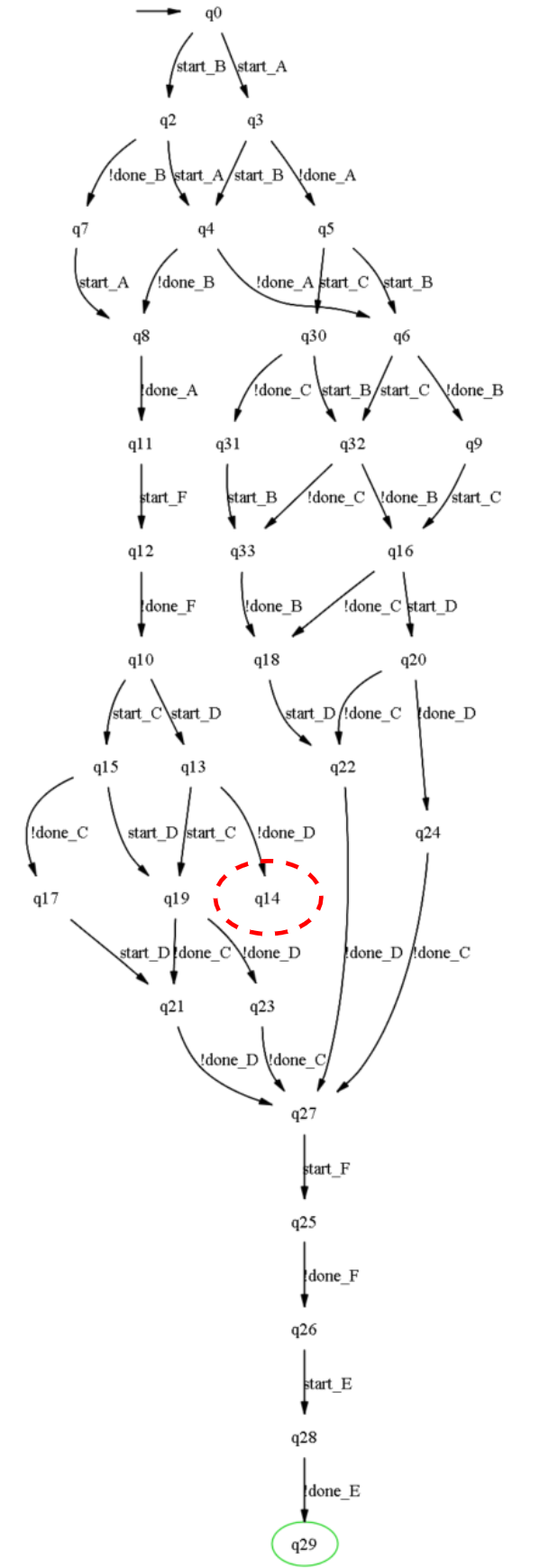}
    \caption{Synchronized system automaton.}
    \label{fig:CompleteAutomaton}
\end{figure}

Using \supremica's \textit{Analyzer} tool, the tasks and constraints were synchronized, and the resulting automaton is shown in Figure~\ref{fig:CompleteAutomaton}. The only marked state, \text{$q_{29}$} highlighted with a green circle at the bottom, 
%%MF which represents the state in which all tasks of the assembly process
represents that all tasks
have been completed. The automaton consists of 33 states and 45 transitions and contains a blocking state, \text{$q_{14}$}, indicated by a dashed red circle. This state is blocking because there are no outgoing transitions from it that respect the defined constraints, making it impossible to reach the marked state.

\begin{figure}[h]
    \centering
    \includegraphics[width=0.2\textwidth]{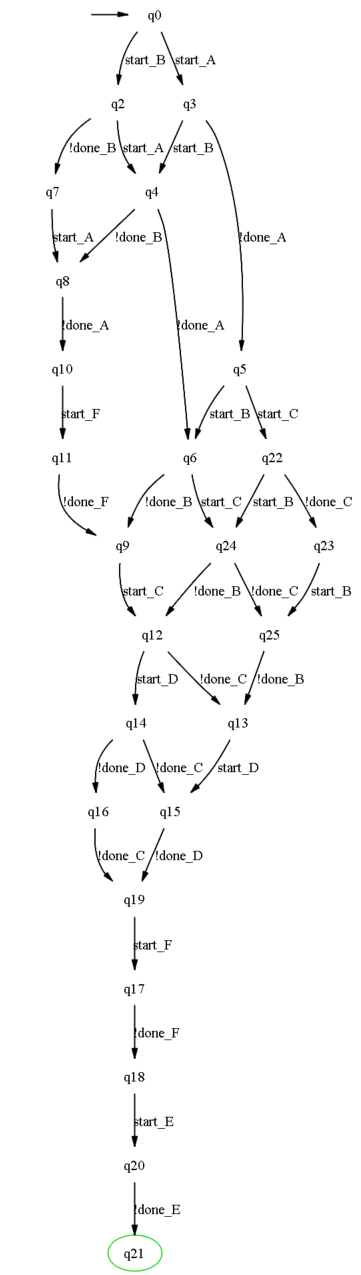}
    \caption{Minimized supervisor automaton.}
    \label{fig:ReducedAutomaton}
\end{figure}

By synthesizing a controllable and non-blocking supervisor using the monolithic algorithm and applying language equivalence minimization, the automaton shown in Figure~\ref{fig:ReducedAutomaton} is obtained. The resulting supervisor consists of 25 states and 34 transitions. The synthesis process disables the blocking state \text{$q_{14}$}. Since the event leading to \text{$q_{14}$} is uncontrollable, the preceding state is also removed. If that state were also uncontrollable, the supervisor would continue removing previous states, tracing back to the first controllable state.

As a result, the synthesized supervisor ensures that the system follows only feasible assembly sequences, guaranteeing successful completion of the assembly process.

For large-scale assembly systems, \supremica's monolithic algorithm may not be able to synthesize a supervisor due to state-space explosion. Alternative synthesis techniques such as BDD-based methods or compositional abstraction-based synthesis~\cite{Malik:JDEDS:2023} would enhance scalability.

%% file: Conclusions.tex
The proposed work enhances the modeling of dependencies and constraints between assembly tasks by incorporating not only static precedence constraints but also more detailed dependencies that reflect various process requirements. This feature meets the needs of modern assembly processes, which require greater flexibility to handle complex dependencies and dynamic production conditions.

Using SCT and synthesizing a controllable, non-blocking, and minimally restrictive supervisor in \textsc{Supremica}, \textit{all} feasible assembly sequences for process completion were generated. These sequences can serve as a basis for further optimization to achieve specific objectives or for developing an exception management system to handle disruptions.

A challenge of this approach is the complexity of modeling certain dynamic constraints, which may require intricate formulations. A potential solution is to express such constraints as Boolean expressions and automatically convert them into finite automata, streamlining the modeling process.

%% file: conference_101719.bbl
\begin{thebibliography}{10}
\providecommand{\url}[1]{#1}
\csname url@samestyle\endcsname
\providecommand{\newblock}{\relax}
\providecommand{\bibinfo}[2]{#2}
\providecommand{\BIBentrySTDinterwordspacing}{\spaceskip=0pt\relax}
\providecommand{\BIBentryALTinterwordstretchfactor}{4}
\providecommand{\BIBentryALTinterwordspacing}{\spaceskip=\fontdimen2\font plus
\BIBentryALTinterwordstretchfactor\fontdimen3\font minus \fontdimen4\font\relax}
\providecommand{\BIBforeignlanguage}[2]{{%
\expandafter\ifx\csname l@#1\endcsname\relax
\typeout{** WARNING: IEEEtran.bst: No hyphenation pattern has been}%
\typeout{** loaded for the language `#1'. Using the pattern for}%
\typeout{** the default language instead.}%
\else
\language=\csname l@#1\endcsname
\fi
#2}}
\providecommand{\BIBdecl}{\relax}
\BIBdecl

\bibitem{assemblychallenges}
D.~Rossit, F.~Tohmé, and M.~Frutos, ``An industry 4.0 approach to assembly line resequencing.'' \emph{Int J Adv Manuf Technol}, vol. 105, p. 3619–3630, 2019.

\bibitem{productvariant}
S.~Hu, J.~Ko, L.~Weyand, H.~ElMaraghy, T.~Lien, Y.~Koren, H.~Bley, G.~Chryssolouris, N.~Nasr, and M.~Shpitalni, ``Assembly system design and operations for product variety,'' \emph{CIRP Annals}, vol.~60, no.~2, pp. 715--733, 2011.

\bibitem{masscustomiz}
F.~S. Fogliatto, G.~J. {da Silveira}, and D.~Borenstein, ``The mass customization decade: An updated review of the literature,'' \emph{International Journal of Production Economics}, vol. 138, no.~1, pp. 14--25, 2012.

\bibitem{asp}
A.~C. Sanderson, L.~S. Homem~de Mello, and H.~Zhang, ``Assembly sequence planning,'' \emph{AI Magazine}, vol.~11, no.~1, p.~62, Mar. 1990.

\bibitem{aspref1}
Z.~Han, Y.~Wang, and D.~Tian, ``Ant colony optimization for assembly sequence planning based on parameters optimization,'' \emph{Frontiers of Mechanical Engineering}, vol.~16, no.~2, pp. 393--409, 2021.

\bibitem{aspref2}
H.~Lv and C.~Lu, ``An assembly sequence planning approach with a discrete particle swarm optimization algorithm,'' \emph{The International Journal of Advanced Manufacturing Technology}, vol.~50, no.~5, pp. 761--770, 2010.

\bibitem{industry5.0}
A.~Adel, ``Future of industry 5.0 in society: human-centric solutions, challenges and prospective research areas,'' \emph{Journal of Cloud Computing}, vol.~11, no.~1, 2022.

\bibitem{des}
C.~G. Cassandras and S.~Lafortune, \emph{Introduction to Discrete Event Systems}, 2nd~ed.\hskip 1em plus 0.5em minus 0.4em\relax Springer Publishing Company, Incorporated, 2010.

\bibitem{sct}
P.~J. Ramadge and W.~M. Wonham, ``Supervisory control of a class of discrete event processes,'' \emph{SIAM Journal on Control and Optimization}, vol.~25, no.~1, pp. 206--230, 1987.

\bibitem{supremica}
K.~Akesson, M.~Fabian, H.~Flordal, and R.~Malik, ``Supremica - an integrated environment for verification, synthesis and simulation of discrete event systems,'' in \emph{2006 8th International Workshop on Discrete Event Systems}, 2006, pp. 384--385.

\bibitem{exsupr}
\BIBentryALTinterwordspacing
B.~Bonafilia, P.~Carlsson, S.~Nilsson, and M.~Fabian, ``Robust manual control of a manufacturing system using supervisory control theory,'' \emph{IFAC Proceedings Volumes}, vol.~47, no.~3, pp. 748--753, 2014, 19th IFAC World Congress. [Online]. Available: \url{https://www.sciencedirect.com/science/article/pii/S1474667016417040}
\BIBentrySTDinterwordspacing

\bibitem{Malik:JDEDS:2023}
\BIBentryALTinterwordspacing
R.~Malik, S.~Mohajerani, and M.~Fabian, ``A survey on compositional algorithms for verification and synthesis in supervisory control,'' \emph{Discrete Event Dynamic Systems}, vol.~33, no.~3, pp. 279--340, Sep 2023. [Online]. Available: \url{https://doi.org/10.1007/s10626-023-00378-8}
\BIBentrySTDinterwordspacing

\end{thebibliography}
